\documentclass{iopart}
\usepackage{euscript,iopams,amssymb,amsfonts,graphicx,bm}
\usepackage{pgfplots,setstack}

\usepackage{float}
\usepackage{epsfig}
\usepackage{esint}
 \usepackage{tikz-cd}

\usepackage{bm,braket}

\eqnobysec

\newcommand{\calU}{{\mathcal U}}
\newcommand{\calG}{{\mathcal G}}
\newcommand{\calGG}{\widetilde{\mathcal G}}
\newcommand{\calH}{{\mathcal H}}
\newcommand{\calHH}{\widetilde{\mathcal H}}
\newcommand{\calP}{{\mathcal P}}

\newcommand{\R}{{\mathbb R}}

\newcommand{\X}{\mathbf{X}}

\renewcommand{\P}{\mathbb{P}}

\newcommand{\x}{\mathbf{x}}
\newcommand{\y}{\mathbf{y}}
\renewcommand{\e}{{\mathrm e}}
\newcommand{\E}{{\mathbb E}}

\newcommand{\n}{\mathbf n}
\newcommand{\calT}{{\mathcal T}}
\renewcommand{\P}{\mathbb P}
\newcommand{\p}{\widetilde{p}}

\newcommand{\G}{\widetilde{G}}
\renewcommand{\H}{\widetilde{H}}
\newcommand{\J}{\widetilde{J}}

\begin{document}


\title[Encounter-based reaction-subdiffusion model I]{Encounter-based reaction-subdiffusion model I: surface adsorption and the local time propagator}

\author{Paul C. Bressloff}
\address{Department of Mathematics, University of Utah 155 South 1400 East, Salt Lake City, UT 84112}

\begin{abstract} 
In this paper, we develop an encounter-based model of partial surface adsorption for fractional diffusion in a bounded domain. We take the probability of adsorption to depend on the amount of particle-surface contact time, as specified by a Brownian functional known as the boundary local time $\ell(t)$.  If the rate of adsorption is state dependent, then the adsorption process is non-Markovian, reflecting the fact that surface activation/deactivation proceeds progressively by repeated particle encounters. The generalized adsorption event is identified as the first time that the local time crosses a randomly generated threshold. Different models of adsorption (Markovian and non-Markovian) then correspond to different choices for the random threshold probability density $\psi(\ell)$. The marginal probability density for particle position $\X(t)$ prior to absorption depends on $\psi$ and the joint probability density for the pair $(\X(t),\ell(t))$, also known as the local time propagator. In the case of normal diffusion one can use a Feynman-Kac formula to derive an evolution equation for the  propagator. Here we derive the local time propagator equation for fractional diffusion by taking a continuum limit of a heavy-tailed continuous-time random walk (CTRW).
We begin by considering a  CTRW on a one-dimensional lattice with a reflecting boundary at $n=0$. We derive an evolution equation for the joint probability density of the particle location $N(t)\in \{n\in {\mathbb Z},n\geq 0\}$ and the amount of time $\chi(t)$ spent at the origin.  The continuum limit involves rescaling $\chi(t)$ by a factor $1/\Delta x$, where $\Delta x$ is the lattice spacing. In the limit $\Delta x \rightarrow 0$, the rescaled functional $\chi(t)$ becomes the Brownian local time at $x=0$. We use our encounter-based model to investigate the effects of subdiffusion and non-Markovian adsorption on the long-time behavior of the first passage time (FPT)  density in a finite interval $[0,L]$ with a reflecting boundary at $x=L$. In particular, we determine how the choice of function $\psi$ affects the large-$t$ power law decay of the FPT density. Finally, we indicate how to extend the model to higher spatial dimensions.

\end{abstract}

\maketitle

\newpage

\section{Introduction}

Encounter-based models of diffusion-mediated surface reactions assume that the probability of adsorption depends upon the amount of particle-surface contact time \cite{Grebenkov20,Grebenkov22,Bressloff22,Bressloff22b,Grebenkov22a}. The latter is determined by a Brownian functional known as the boundary local time $\ell(t)$ \cite{Ito63,McKean75,Majumdar05}. It has subsequently been shown that encounter-based models can also be applied to non-diffusive processes such as active particles \cite{Bressloff22rtp,Bressloff23,Bressloff23a}, where the particle-surface contact time is the amount of time the particle spends ``stuck'' to the boundary, and to diffusion in domains with partially absorbing interior traps \cite{Bressloff22,Bressloff22a,Bressloff22c}. In the latter case, a particle freely enters and exits a trap, but can only be absorbed within the trapping region. The probability of adsorption is taken to depend on the particle-trap encounter time, which is given by the Brownian occupation time. 

There are two basic components of encounter-based models that underscore their general applicability:  (i) The stochastic process of adsorption at a surface (or absorption within an interior trap) is separated from the bulk dynamics. This means that the probability of adsorption can be taken to depend on the particle-surface contact time.  If the rate of adsorption is state dependent, then the adsorption process is non-Markovian, reflecting the fact that surface activation/deactivation proceeds progressively by repeated particle encounters \cite{Bartholomew01,Filoche08}. Alternatively, adsorption could involve the exit of a particle through a stochastically-gated ion channel or pore, which may require multiple return visits to the channel before it is open \cite{Bressloff15}. (ii) The generalized adsorption event is identified as the first time that the particle-surface contact time crosses a randomly generated threshold. Different models of adsorption (Markovian and non-Markovian) then correspond to different choices for the random threshold probability density $\psi$. In order to incorporate this form of adsorption, it is necessary to determine the joint probability density or generalized propagator for particle position and the boundary local time. This can be achieved by solving a classical boundary value problem (BVP) for the probability density of particle position and a constant rate of adsorption. (In the case of normal diffusion, this takes the form of  a Robin or radiation BVP.)  The constant adsorption rate is then reinterpreted as a Laplace variable $z$ conjugate to the local time, and the inverse Laplace transform of the classical solution with respect to $z$ yields the propagator.

In this paper, we develop an encounter-based model of partial surface adsorption for a fractional diffusion equation based on the continuum limit of a continuous-time random walk (CTRW) \cite{Hughes95}. The latter are widely used in studies of trapping-based mechanisms for anomalous diffusion \cite{Barkai00,Metzler04}. For example, there are several mechanisms of intracellular transport that involve the transient trapping of diffusing particles, resulting in anomalous subdiffusion on intermediate timescales and normal diffusion on long timescales. However, in the case of an infinite hierarchy of arbitrarily deep but rare traps, anomalous subdiffusion can occur at all times. It is this type of process that is modeled in terms of a CTRW. The basic idea is that trapping increases the time between jumps (waiting times) of a classical random walk. This means that the classical exponential waiting time distribution, which is equivalent to taking constant hopping rates in the associated master equation, is replaced by a heavy-tailed waiting time distribution. One of the interesting consequences of heavy tails is that the resulting CTRW is weakly non-ergodic; the temporal average of a long particle trajectory differs from the ensemble average over many diffusing particles \cite{He08,Weigel11,Burov11,Jeon11,Jeon12,Metzler14}. 
Here we focus on the joint effects of subdiffusion and non-Markovian adsorption on the first passage time (FPT) density.

The structure of the paper is as follows. In section 2 we construct the local time propagator equation for fractional diffusion on the half-line. In order to introduce the encounter-based method, we first briefly summarize the case of normal diffusion.  We then derive the propagator equation for a CTRW by combining the analysis of Feynman-Kac equations for functionals of CTRWs \cite{Carmi11} with the analysis of a CTRW with a reactive boundary \cite{Metzler07}. The corresponding particle-boundary contact time is simply the amount of time spent at the lattice site $n=0$, which is given by the functional $\chi(t)=\int_0^t\delta_{N(\tau),0} d\tau$ where $N(\tau)$ is the lattice site occupied at time $\tau$. We then show how to obtain the local time propagator equation for fractional diffusion by taking a continuum limit of a heavy-tailed CTRW. The continuum limit involves rescaling $\chi(t)$ by a factor $1/\Delta x$, where $\Delta x$ is the lattice spacing. In the limit $\Delta x \rightarrow 0$, the rescaled functional $\chi(t)$ can be formally identified as the Brownian local time.
In section 3 we consider the first passage time (FPT) density for fractional diffusion in a finite interval $[0,L]$ with a partially absorbing boundary at $x=0$ and a totally reflecting boundary at $x=L$. Since the moments of the FPT density are infinite for a subdiffusive process, we focus instead on the long-time  behavior of the FPT density. The latter can be extracted from the small-$s$ behavior of the corresponding Laplace transformed FPT density. We show how the FPT density can be expanded as an asymptotic series in fractional powers of the time $t$. In particular, if the local time threshold density $\psi(\ell)$ has finite moments, then the $n$-th term in the asymptotic expansion, $n\geq 1$, is proportional to $\tau_nt^{-n\alpha-1}$, where $0<\alpha<1$ and $\tau_n$ is the $n$-th moment of the FPT in the case of normal diffusion ($\alpha=1$). This generalizes  previous results that were obtained for the fractional diffusion equation with Dirichlet or Robin boundary conditions (constant rate of adsorption) \cite{Condamin07,Yuste07,Grebenkov10}. We also consider a few examples of heavy-tailed distributions for $\psi(\ell)$. In these cases the dominant power law decay of the FPT density at large times $t$ has contributions from two distinct anomalous processes: subdiffusion within the bulk domain and the random threshold for adsorption at the boundary. Finally, in section 4, we indicate how to extend the analysis to higher spatial dimensions. In a companion paper \cite{PCBII}, we develop a corresponding theory for reaction-subdiffusion in the presence of a partially absorbing trapping domain, which is based on the construction of a fractional diffusion equation for the occupation time propagator.

\section{Encounter-based reaction-subdiffusion model on the half-line}

\subsection{Normal diffusion} In order to motivate the encounter-based model of fractional diffusion, it is useful to briefly recall the corresponding model for normal diffusion \cite{Grebenkov20,Bressloff22}. 
Consider a particle diffusing in the half-line $x\in [0,\infty)$. First suppose that the boundary at $x=0$ is totally reflecting. Introduce the boundary local time
\begin{equation}
\label{1Dloc}
\ell(t)=\lim_{\epsilon\rightarrow 0} \frac{D}{\epsilon} \int_0^tH(\epsilon-X(\tau)d\tau=D\int_0^{t} \delta(X(\tau))d\tau,
\end{equation}
where $H(x)$ is the Heaviside function. Let $P(x,\ell,t)$ denote the local time propagator, which is defined to be the joint probability density for particle position $X(t)$ and $\ell(t)$. Using a Feynman-Kac equation, it can be shown that the propagator evolves according to
\numparts
\begin{eqnarray}
\label{norma}
	&\frac{\partial P(x,\ell,t)}{\partial t} =D\frac{\partial^2P(x,\ell,t)}{\partial x^2} \mbox{ for } x \in  (0,\infty),\\
&\left . \frac{\partial P(x,\ell,t) }{\partial x}\right |_{x=0}=\frac{\partial P}{\partial \ell}(0,\ell,t)  +\delta(\ell)P(0,0,t) .
\label{normb} 
	\end{eqnarray}
	\endnumparts	
	Now suppose that the boundary at $x=0$ is partially absorbing. Furthermore, assume that the probability of adsorption depends on the amount of contact time between the particle and the boundary, which is determined by $\ell(t)$.
Introduce the stopping time 
\begin{equation}
\label{1DTell}
{\mathcal T}=\inf\{t>0:\ \ell(t) >\widehat{\ell}\},
\end{equation}
where $\widehat{\ell}$ is a random threshold with $\P[\widehat{\ell}>\ell]=\Psi(\ell)$. The stopping time ${\mathcal T}$ is the FPT for the event that $\ell(t)$ crosses the random threshold $\widehat{\ell}$, which we identify with the time at which adsorption occurs.  For a given distribution $\Psi$, let $p^{\Psi}(x,t)$ denote the marginal probability density at time $t$:
\[p^{\Psi}(x,t)dx=\P[x<X(t)<x+dx, \ t < {\mathcal T}].\]
Since $\ell(t)$ is a nondecreasing process, the condition $t < {\mathcal T}$ is equivalent to the condition $\ell(t) <\widehat{\ell}$. 
This implies 
\begin{eqnarray}
\fl p^{\Psi}(x,t)&=\int_0^{\infty} d\ell \ \psi(\ell)\int_0^{\ell} d\ell' P(x,\ell',t)=\int_0^{\infty} d\ell' P(x,\ell',t)\int_{\ell'}^{\infty}  \psi(\ell)d\ell \nonumber \\
\fl &=\int_0^{\infty}\Psi(\ell) P (x,\ell,t)d\ell .
\label{1DpPsi}
\end{eqnarray}
The penultimate line follows from reversing the order of integration, and $\psi(\ell)=-\Psi'(\ell)$ is the probability density of the random threshold $\widehat{\ell}$. First suppose that $\widehat{\ell}$ is exponentially distributed so that $\Psi(\ell)=\e^{-z \ell}$ for constant $z$. Equation (\ref{1DpPsi}) implies that $p(x,t)$ (after dropping the superscript $\Psi$) is the Laplace transform of the propagator with respect to $\ell$:
\begin{equation}
p(x,t)=\int_0^{\infty}\e^{-z\ell}P(x,\ell,t)d\ell :=G(x,z,t),
\end{equation}
with 
\numparts
\begin{eqnarray}
\label{1Dalpha}
&\frac{\partial G(x,z,t)}{\partial t}=D\frac{\partial^2 G(x,z,t)}{\partial x^2},\quad x>0,\\
\label{1Dalphb}
& \frac{\partial G(0,z,t)}{\partial x}  =  zG(0,z,t)  . 
\end{eqnarray}
\endnumparts 
Note that the generator $G(x,z,t)$ satisfies the classical diffusion equation with a Robin boundary condition at $x=0$, and can be solved using standard methods. 
Finally, given $G(x,z,t)$, the density $p^{\Psi}(x,t)$ for non-exponential $\Psi$ can be obtained by inverting the solution with respect to $z$:
\begin{eqnarray}
     \label{1Dint}
  p^{\Psi} (x,t)&=\int_0^{\infty} \Psi(\ell){\mathcal L}_{\ell}^{-1}[G(x,z,t)]d\ell ,
  \end{eqnarray}
  where ${\mathcal L}^{-1}$ denotes the inverse Laplace transform.

The local time does not change when the particle is diffusing in the bulk, which suggests that the local time propagator equation for a fractional diffusion process could be obtained by replacing $D$ with a fractional differential operator. For example, in the case of a subdiffusive process,
\numparts
\begin{eqnarray}
\label{fracPa}
	&\frac{\partial P(x,\ell,t)}{\partial t} =K_{\alpha}{\mathcal D}_t^{1-\alpha}\frac{\partial^2P(x,\ell,t)}{\partial x^2} \mbox{ for } x \in  (0,\infty),\\
&\left . \frac{\partial P(x,\ell,t) }{\partial x}\right |_{x=0}=\frac{\partial P}{\partial \ell}(0,\ell,t)  +\delta(\ell)P(0,0,t)  ,
\label{fracPb}
	\end{eqnarray}
	\endnumparts	
where the fractional derivative $ {\mathcal D}_t^{1-\alpha}  $ is defined in Laplace space according to \cite{Barkai00}
\begin{equation}
\int_0^{\infty}\e^{-st}  {\mathcal D}_t^{1-\alpha} f(t)dt=s^{1-\alpha} \widetilde{f}(s).
\end{equation}
It can also be written as the fractional Riemann-Liouville derivative \cite{Barkai00}
\begin{equation}
 {\mathcal D}_t^{1-\alpha} f(t)=\frac{1}{\Gamma(a)}\frac{\partial}{\partial t}\int_0^t\frac{f(t')}{(t-t')^{1-\alpha}}dt'.
 \end{equation}
 where $\Gamma(\alpha)$ is the gamma function. One could then proceed by Laplace transforming with respect to $\ell$ to obtain a fractional diffusion equation for the generator with a Robin boundary condition at $x=0$:
 \numparts
\begin{eqnarray}
\label{alpha}
&\frac{\partial G(x,z,t)}{\partial t}=K_{\alpha} {\mathcal D}_t^{1-\alpha}\frac{\partial^2 G(x,z,t)}{\partial x^2},\quad x>0,\\
\label{alphb}
& \frac{\partial G(0,z,t)}{\partial x}  =  zG(0,z,t)  . 
\end{eqnarray}
\endnumparts 
(This type of equation has been analyzed in Ref. \cite{Grebenkov10} for finite $z$ and in Ref. \cite{Yuste07} for $z\rightarrow \infty$.) The corresponding marginal probability density for a general distribution $\Psi(\ell)$ is then determined by substituting the solution of equations (\ref{alpha}) and (\ref{alphb}) into equation (\ref{1Dint}).

However, as we highlighted in the introduction, a more principled way of deriving a fractional diffusion equation for a subdiffusive process is to consider the continuum limit of a CTRW. Therefore, in this section we show how equations (\ref{fracPa}) and (\ref{fracPb}) follow from taking the continuum limit of a corresponding propagator for a heavy-tailed CTRW. In particular, we find that the derivation of the boundary condition (\ref{1Dalphb}) is non-trivial.
 
 \subsection{Propagator equation for a CTRW}
 
Consider a CTRW on a 1D lattice $\{n, n\geq 0\}$ and a reflecting boundary at $n=0$. Let $N(t)$ be the lattice site occupied at time $t$. Waiting times between jump events are independent identically distributed random variables with probability density $u(\tau)$. For simplicity, we assume that the CTRW is unbiased so that for all $n\geq 1$, the jumps $n\rightarrow n\pm 1$ occur with probability 1/2. Given the stochastic process $N(t)$, define the functional
\begin{equation}
\label{Ldef}
\chi(t)=\int_0^tF(N(\tau))d\tau,\quad F(N(\tau))=\overline{h} \delta_{N(\tau),0},
\end{equation}
with $\overline{h}$ a positive constant. (In section 2,3 we will show that for an appropriate choice of $\overline{h}$, $\chi(t)\rightarrow \ell(t)$ in the continuum limit, where $\ell(t)$ is the boundary local time). Note that $\chi(t)$ is a positive, non-decreasing function of time.  Let $P_n(\ell,t)$ denote the joint probability density or propagator for the pair $(N(t),\chi(t))$. It follows that
 \begin{eqnarray}
 \label{A1}
& P_n(\ell,t)=\bigg \langle \delta\left (\ell -\chi(t) \right )\bigg \rangle_{N(0)=n_0}^{N(t)=n} ,
 \end{eqnarray}
 where expectation is taken with respect to all CTRWs realized by $N(\tau)$ between $N(0)=n_0$ and $N(t)=n$. 
  Introduce the generator
 \begin{eqnarray}
G_n(z,t)&=\bigg\langle \exp \left ( -z \chi(t)\right )\bigg \rangle_{N(0)=n_0}^{N(t)=n}=\int_0^{\infty} \e^{-z\ell}P_n(\ell,t)d\ell.
 \end{eqnarray}
Analogous to Brownian functionals, one can derive an evolution equation for the propagator using a discrete version of a Feynman-Kac formula. This was previously shown for an infinite lattice in Ref. \cite{Carmi11}. Here we derive the corresponding formula in the case of a reflecting boundary at $n=0$ by adapting a study of a CTRW with a reactive boundary \cite{Metzler07}.

Let $w_n(\ell,t)dt$ be the probability that the particle jumps to the state $N(t)=n$ and $\chi(t)=\ell$ in the time interval $[t,t+dt]$. The propagator away from the boundary can be expressed as
\begin{equation}
\label{Pw0}
P_n(\ell,t)=\int_0^tU(\tau)w_n(\ell-\tau F(n),t-\tau)d\tau, \quad n \geq 1,
\end{equation}
where $U(\tau)=1-\int_0^{\tau}u(\tau')d\tau'$ is the probability of not jumping in a time interval of length $\tau$. That is, if the last jump was at time $t-\tau$ then over the time interval $[t-\tau,t]$ we simply have $\Delta \ell=\tau F(n)$. The next step is to derive a recursion relation for $w_n$ by noting that to arrive at $(n,\ell)$ at time $t$, the particle must have hopped from one of the neighboring sites $n\pm 1$ for all $n\geq 1$. Assuming that the last jump occurred at time $t-\tau$ and the CTRW is unbiased, we have
\begin{eqnarray}
w_n(\ell,t)&=P_n^{(0)} \delta(\ell)\delta(t)+\frac{1}{2}\int_0^t u(\tau)w_{n+1}(\ell-\tau F(n+1),t-\tau)d\tau\nonumber \\
&\quad +\frac{1}{2}\int_0^t u(\tau)w_{n-1}(\ell-\tau F(n-1),t-\tau)d\tau 
\end{eqnarray}
for all $n\geq 1$ with
$P^{(0)}_n=P_n(\ell=0,t=0)$.
Laplace transforming with respect to $\ell$ by setting $H_n(z,t)=\int_0^{\infty}\e^{-z\ell}w_n(\ell,t)d\ell$ leads to the equation
\begin{eqnarray}
H_n(z,t)&=P^{(0)}_n \delta(t) +\frac{1}{2}\int_0^t u(\tau)\e^{-z\tau F(n+1)}H_{n+1}(z,t-\tau)d\tau\nonumber \\
&\quad +\frac{1}{2}\int_0^t u(\tau)\e^{-z\tau F(n-1)}H_{n-1}(z,t-\tau)d\tau,
\end{eqnarray}
and Laplace transforming the result with respect to $s$ yields
\begin{eqnarray}
\label{wn}
\H_n(z,s)&=P^{(0)}_n +\frac{1}{2}  \widetilde{u}(s+z  F(n+1))\H_{n+1}(z,s) \nonumber \\
&\quad +\frac{1}{2}\widetilde{u}(s+z  F(n-1))\H_{n-1}(z,s),
\end{eqnarray}
where $\widetilde{u}(s)=\int_0^{\infty}\e^{-s\tau} u(\tau)d\tau$. Denoting the propagator at the boundary by
\begin{equation}
\label{Pw}
\calP_0(\ell,t)=\int_0^tU(\tau)w_0(\ell-\tau F(0),t-\tau)d\tau, \quad n \geq 1,
\end{equation}
and setting $\calG_0(z,t)=\int_0^{\infty}\e^{-z\ell}\calP_0(\ell,t)d\ell$ and $\calH_0(z,t)=\int_0^{\infty}\e^{-z\ell}w_0(\ell,t)d\ell$, we find that
\begin{eqnarray}
\label{w0}
\fl \calHH_0(z,s)&=\calP^{(0)}_0 +\frac{1}{2}  \widetilde{u}(s+z  F(1))\H_{1}(z,s)  +\frac{1}{2}\widetilde{u}(s+z  F(0))\calHH_{0}(z,s).
\end{eqnarray}
This follows from introducing an auxiliary site at $n=-1$, which is a common method for treating reflecting boundary conditions in finite difference schemes.

Performing the double Laplace transform of equation (\ref{Pw}) shows that
\numparts
\begin{eqnarray}
\G_n(z,s)&=\widetilde{U}(s+zF(n)) \H_n(z,s),\\
 \calGG_0(z,s)&=\widetilde{U}(s+zF(0)) \calHH_0(z,s),
\end{eqnarray}
\endnumparts
with $\widetilde{U}(s)=(1-\widetilde{u}(s))/{s}$.
Substituting into equation (\ref{wn}) leads to the recursion equations  
\begin{eqnarray}
\label{Pn}
\fl &(s+zF(n))\G_n(z,s)=P^{(0)}_n +\frac{1}{2}  \bigg \{\widetilde{v}(s+z  F(n+1))\G_{n+1}(z,s) \nonumber \\
\fl &\hspace{2cm} + \widetilde{v}(s+z  F(n-1))\G_{n-1}(z,s)-2 \widetilde{v}(s+z  F(n))\G_{n}(z,s)\bigg \},\ n\geq 1,
\end{eqnarray}
where $\widetilde{v}(s)={s\widetilde{u}(s)}/(1-\widetilde{u}(s))$
and, at the boundary,
\begin{eqnarray}
\label{P0}
\fl  &(s+zF(0))\calGG_0(z,s)=\calP^{(0)}_0 +\frac{1}{2}\bigg \{\widetilde{v}(s+z  F(1))\G_{1}(z,s)- \widetilde{v}(s+z  F(0))\G_{0}(z,s)\bigg \}. 
\end{eqnarray}
Finally, setting $F(n)=\overline{h}\delta_{n,0}$ gives
\numparts
\begin{eqnarray}
\label{gena}
\fl  &s\G_n(z,s)=P^{(0)}_n  +\frac{\widetilde{v}(s)}{2}  \bigg \{ \G_{n+1}(z,s)  +\G_{n-1}(z,s)-2 \G_{n}(z,s)\bigg \},\ n\geq 1,\\
 \fl & (s+z\overline{h})\frac{\widetilde{v}(s)}{\widetilde{v}(s+z\overline{h})} \G_0(z,s)=\calP^{(0)}_0  +\frac{\widetilde{v}(s)}{2} \bigg\{\G_{1}(z,s)  -\widetilde{G}_0(z,s) \bigg\},
 \label{genb}
 \end{eqnarray}
\endnumparts
where we have set
\begin{equation}
\label{ggen}
\widetilde{G}_0(z,s) =\frac{\widetilde{v}(s+z\overline{h})}{\widetilde{v}(s)}\calGG_{0}(z,s).
\end{equation}

\subsection{Continuum limit}

In the special case of an exponential waiting-time density, $u(\tau)=h\e^{-h\tau}$, the relevant Laplace transforms are
$\widetilde{u}(s) ={h}/(h+s)$ and $\widetilde{v}(s)=h$.
Equations (\ref{gena}) and (\ref{genb}) then reduce to the form
  \begin{eqnarray}
 \fl s \G_n(z,s)=\frac{h}{2}[\G_{n+1}(z,s)+\G_{n-1}(z,s)-2\G_n(z,s] -z \overline{h}\G_0(z,s) \delta_{n,0}
 \label{calG}
\end{eqnarray}
for $n\geq 0$, and we have set $\G_{-1}=\G_{0}$ for notational convenience. Inverting with respect to $z$ and $s$ leads to the Feynman-Kac equation for a classical unbiased random walk with a constant hopping rate $h$:
\begin{eqnarray}
\frac{\partial P_n(\ell,t)}{\partial t}&=\frac{h}{2}[P_{n+1}(\ell,t)+P_{n-1}(\ell,t) -2P_{n+1}(\ell,t)] \nonumber\\
&\quad -\overline{h}\left[\frac{\partial P_0}{\partial \ell}(\ell,t) + \delta(\ell)P_0(0,t)\right ]\delta_{n,0},
\label{calP}
\end{eqnarray}
with $P_{-1}=P_0$. In order to take a continuum limit of equation (\ref{calP}), we set
\begin{equation}
\label{hhbar}
h=\frac{2D}{\Delta x^2},\quad \overline{h} =\frac{D}{\Delta x},
\end{equation}
so that
\begin{equation}
L(t)=\frac{D}{\Delta x}\int_0^t\delta_{N(\tau),0}d\tau .
\end{equation}
In the continuum limit, $L(t)\rightarrow \ell(t)$ with
\begin{equation}
\ell(t)=\lim_{\Delta x\rightarrow 0} \frac{D}{\Delta x}\int_0^t \delta_{X(\tau),0}d\tau =D\int_0^t\delta(X(\tau))d\tau.
\label{loc}
\end{equation}
This a formal definition of the Brownian local time \cite{Ito63,McKean75,Majumdar05} scaled by the diffusivity $D$ for convenience. 
Substituting for $h$ and $\overline{h}$ in the propagator  equation  (\ref{calP}) gives
\begin{eqnarray}
  \frac{\partial P_n(\ell,t)}{\partial t}&=D\frac{P_{n+1}(\ell,t)+P_{n-1}(\ell,t) -2P_{n}(\ell,t)]}{\Delta x^2} \nonumber\\
  &\quad -\frac{D}{\Delta x}\frac{\partial P_0}{\partial \ell}(\ell,t)- \frac{D}{\Delta x}\delta(\ell)P_0(0,t),
\label{calP2}
\end{eqnarray}
with $P_{-1}=P_0$.
It follows that
\numparts
\begin{eqnarray}
 \fl\frac{\partial P_n(\ell,t)}{\partial t}&=D\frac{P_{n+1}(\ell,t)+P_{n-1}(\ell,t) -2P_{n}(\ell,t)]}{\Delta x^2},\quad n \geq 1,\\
\fl \Delta x\frac{\partial P_0(\ell,t)}{\partial t}&=D\frac{P_{1}(\ell,t)-P_{0}(\ell,t)}{\Delta x}-D\left [\frac{\partial P_0}{\partial \ell}(\ell,t)  +\delta(\ell)P_0(0,t)\right ].\end{eqnarray}
\endnumparts
Taking the limit $\Delta x\rightarrow 0$ with $P(x=n\Delta x,\ell,t)\Delta x=P_n(\ell,t)$ then recovers the local time  propagator equation for reflected BM on the half-line given by equations (\ref{norma}) and (\ref{normb}).

Obtaining a continuum limit of the general CTRW propagator equation, see equations (\ref{gena}) and (\ref{genb}), is more involved.
Following Refs.  \cite{Metzler07,Carmi11}, consider the heavy-tailed waiting time density
\begin{equation}
u(\tau)\sim \frac{B_{\alpha}}{|\Gamma(-\alpha)|}\tau^{-(1+\alpha)},\quad 0 < \alpha <1 .
\end{equation}
The Laplace transform  for small $s$ is then
\begin{equation}
\widetilde{u}(s)\sim 1-B_{\alpha}s^{\alpha} \mbox { and } \widetilde{v}(s)\sim \frac{s}{B_{\alpha}}(s^{-\alpha}-B_{\alpha}).
\end{equation}
Here $B_{\alpha}$ plays the role of the inverse of the hopping rate $h$. We also assume that $u(t)\sim h\e^{-ht}$ for small $t$, so that $\widetilde{u}(s)\sim h/s$ and $\widetilde{v}(s)\sim h$ in the limit $s\rightarrow \infty$.
Given the lattice spacing $\Delta x$, we take $h$ and $\overline{h}$ to be given by equations (\ref{hhbar}), whereas
\begin{equation}
B_{\alpha}=\frac{(\Delta x)^2}{2K_{\alpha}}
\end{equation}
for a constant $K_{\alpha}$. Setting $\G(n\Delta x,z,s)\Delta x=\G_n(z,s)$, $n\geq 0$. 
and taking the limit $\Delta x\rightarrow 0$ in equation (\ref{gena}) gives
\numparts
\begin{eqnarray}
\label{Gx0}
s\G(x,z,s)&=P(x,0,0)+K_{\alpha}s^{1-\alpha}\frac{\partial^2 \G(x,z,s)}{\partial x^2}
\end{eqnarray}
for all $x>0$.
In order to determine the continuum limit of equation (\ref{genb}) we use the fact that $\overline{h}\rightarrow \infty$ as $\Delta x\rightarrow 0$, which implies that $\widetilde{v}(s+z\overline{h})\sim h$ for large $s$. Assuming that $\calP_0^{(0)}\rightarrow 0$ as $\Delta x\rightarrow \infty$, we find that
\begin{eqnarray}
\frac{\partial \G(0,z,s)}{\partial x}&= z  \G(0,z,s).
\label{Gx1}
\end{eqnarray}
\endnumparts
Finally, inverting the Laplace transforms in $s$ and $z$, we obtain the fractional diffusion equation for the local time propagator given by equations (\ref{fracPa}) and (\ref{fracPb}).

\section{FPT problem for reaction-subdiffusion in an interval}

In this section we apply the encounter-based reaction-diffusion model to a FPT problem. In the case of a subdiffusive process, the MFPT for adsorption is infinite even in the case of a bounded domain. However, it is possible to investigate the long-time and short-time asymptotic behavior of the FPT density by considering, respectively, the small-$s$ and large-$s$ behavior of the corresponding Laplace transform. (Since we are dealing with a partially absorbing boundary at $x=0$, it is important to distinguish between the FPT for the particle to reach the boundary for the first time and the FPT for the particle to reach the boundary and be permanently absorbed. That is, the particle may visit $x=0$ and return to the bulk domain multiple times before being absorbed. Hence, the FPT for adsorption is sometimes referred to as the last passage time for exiting the domain.)

\subsection{Derivation of the FPT density} Consider the fractional diffusion equation in a bounded domain $[0,L]$ with a partially absorbing boundary at $x=0$ and a totally reflecting boundary at $x=L$. Furthermore, suppose that the initial condition for the local time propagator is $P(x,\ell,0)=\delta(x-x_0)\delta(\ell)$ for $0<x_0<L$, which implies that $p^{\Psi}(x,0)=\delta(x-x_0)$ in the case of the marginal density defined by equation (\ref{1DpPsi}). 
Introduce the survival probability
\begin{equation}
S^{\Psi}(t)=\int_0^Lp^{\Psi}(x,t)dx .
\end{equation}
Differentiating both sides with respect to $t$ and using equations (\ref{fracPa}) and (\ref{fracPb}) gives
\begin{eqnarray}
\frac{dS^{\Psi}(t)}{dt}&=\int_0^{\infty} \left [\int_0^{\infty} \Psi(\ell)\frac{\partial P(x,\ell,t)}{\partial t}d\ell \right ]dx\nonumber \\
&=\int_0^{\infty} \Psi(\ell)K_{\alpha} {\mathcal D}_t^{1-\alpha}\left [ \int_0^{\infty} \frac{\partial^2 P(x,\ell,t)}{\partial x^2}dx\right ]d\ell\nonumber \\
&=-\int_0^{\infty} \Psi(\ell)\left [K_{\alpha} {\mathcal D}_t^{1-\alpha}  \left . \frac{\partial P(x,\ell,t)}{\partial x}\right |_{x=0}\right ] d\ell\nonumber\\
&=-\int_0^{\infty} \Psi(\ell) K_{\alpha} {\mathcal D}_t^{1-\alpha} \left [ \frac{\partial P(0,\ell,t) }{\partial \ell} +\delta(\ell)P(0,0,t) \right ] d\ell\nonumber\\
&=-K_{\alpha}\int_0^{\infty} \psi(\ell)  {\mathcal D}_t^{1-\alpha}   P(0,\ell,t) d\ell:=-J^{\Psi}(t).
\end{eqnarray}
We have assumed that the order of integration and differentiation can be reversed, and have performed an integration by parts with respect to $\ell$. The term $J^{\Psi}(t)$ is the probability flux into the boundary at $x=0$, and is equivalent to the FPT density for adsorption at $x=0$.

In general it is difficult to obtain an analytical expression for $J^{\Psi}(t)$. Therefore, we will proceed by calculating the Laplace transformed flux
\begin{equation}
\label{jag}
\J^{\Psi}(s)=K_{\alpha}s^{1-\alpha}\int_0^{\infty} \psi(\ell)  {\mathcal L}^{-1}_{\ell} [ \G(0,z,s) ]d\ell,
\end{equation}
with $\G(x,z,s)$ satisfying the BVP
\numparts
\begin{eqnarray}
\label{JGx0}
&K_{\alpha}s^{1-\alpha}\frac{\partial^2 \G(x,z,s)}{\partial x^2}-s\G(x,z,s)&=-\delta(x-x_0),\\
&\frac{\partial \G(0,z,s)}{\partial x}= z  \G(0,z,s).
\label{JGx1}
\end{eqnarray}
\endnumparts
A solution of equation (\ref{JGx0}) for $0\leq x<x_0$ that satisfies the boundary condition at $x=0$ takes the form
\numparts
\begin{eqnarray}
\label{pl}
\G_<(x,z,s)
 &=\sqrt{s^{\alpha}/K_{\alpha}}\cosh(\sqrt{s^{\alpha}/K_{\alpha}}x)+z\sinh(\sqrt{s^{\alpha}/K_{\alpha}}x).
\end{eqnarray}
Similarly, for $L>x>x_0$ with a reflecting boundary condition at $x=L$, we have
\begin{equation}
\label{pg}
\G_>(x,z,s)=\cosh(\sqrt{s^{\alpha}/K_{\alpha} }(L-x)).\end{equation}
\endnumparts
Imposing continuity of the solution across $x=x_0$ and the flux discontinuity condition $ \partial_x\G(x_0^+,z,s)- \partial_x\G(x_0^-,z,s)=-1/(s^{1-\alpha}K_{\alpha})$ implies that
\begin{equation}
\G(x,z,s)=\left \{ \begin{array}{cc} A(z,s)\p_<(x,s)\p_>(x_0,s),& 0\leq x < x_0 \\
A(z,s)\p_<(x_0,s)\p_>(x,s),& x< x_0 \leq L, \end{array} \right .
\label{psticky}
\end{equation}
with
\begin{eqnarray}
 A(z,s)&=\frac{1}{s}\frac{\sqrt{s^{\alpha}/K_{\alpha}}}{\sqrt{s^{\alpha}/K_{\alpha}}\sinh(\sqrt{s^{\alpha}/K_{\alpha}}L)+z\cosh(\sqrt{s^{\alpha}/K_{\alpha}}L)}. 
\end{eqnarray}
Note that $\sqrt{s^{\alpha}/K_{\alpha}}$ has units of inverse length.
Setting $x=0$ in the solution (\ref{psticky}) gives
\begin{eqnarray}
\label{px0}
&&\G(0,z,s)=A(z,s)\sqrt{s^{\alpha}/K_{\alpha}} \cosh(\sqrt{s^{\alpha}/K_{\alpha}}(L-x_0)). 
\end{eqnarray}

Since $A(z,s)$ has a simple pole with respect to $z$, it follows that
\begin{eqnarray}
\fl {\mathcal L}^{-1}_{\ell}\G(0,z,s)
 &=\frac{\cosh(\sqrt{s^{\alpha}/K_{\alpha}}(L-x_0))}{s \cosh(\sqrt{s^{\alpha}/K_{\alpha}}L)}{\mathcal L}^{-1}_{\ell} \left [\frac{s^{\alpha}/K_{\alpha}}{z+\sqrt{s^{\alpha}/K_{\alpha}}\tanh(\sqrt{s^{\alpha}/K_{\alpha}}L)}\right ]\nonumber \\
\fl  &=\frac{\cosh(\sqrt{s^{\alpha}/K_{\alpha}}(L-x_0))}{K_{\alpha}s^{1-\alpha} \cosh(\sqrt{s^{\alpha}/K_{\alpha}}L)}\exp\left (-\sqrt{s^{\alpha}/K_{\alpha}}\tanh(\sqrt{s^{\alpha}/K_{\alpha}}L)\ell\right ).
\label{PPS}
\end{eqnarray}
Plugging in the solution for ${\mathcal L}^{-1}_{\ell}\G(0,z,s)$ into equation  (\ref{jag}) gives
\begin{eqnarray}
\label{JagPsi}
  \J^{\Psi}(s) = \frac{\cosh(\sqrt{s^{\alpha}/K_{\alpha}}(L-x_0))}{\cosh(\sqrt{s^{\alpha}/K_{\alpha}}L)}\widetilde{\psi}  (\Gamma(\sqrt{s^{\alpha}/K_{\alpha}})),
 \end{eqnarray}
 where $\widetilde{\psi}(z)$ is the Laplace transform of $\psi(\ell)$, and
 \begin{equation}
 \label{gamy}
 \Gamma(y)=y\tanh(yL).
 \end{equation}
 Hence, mathematically speaking, $\J(s)$ for an exponential distribution can be mapped to $\J^{\Psi}(s)$ for a non-exponential distribution by taking $s/D\rightarrow s^{\alpha}/K_{\alpha}$. This also follows from the structure of the modified Helmholtz equation (\ref{JGx0}), see Ref. \cite{Grebenkov10}. 
 In general, it is difficult to derive an analytical expression for $J^{\Psi}(t)$ by finding the inverse Laplace transform of equation (\ref{JagPsi}) with respect to the Laplace variable $s$.
 In the special case of an exponential density $\psi(\ell)=\kappa\e^{-\kappa \ell}$, it is possible to derive an infinite series representation of the exact FPT density using a spectral decomposition of the solution to the Robin BVP given by equations (\ref{alpha}) and (\ref{alphb}) \cite{Grebenkov10}. On the other hand, for non-exponential densities $\psi(\ell)$, there is not a simple analog of a Robin boundary condition. Therefore, we will focus on the long-time and short-time behavior of $J^{\Psi}(t)$ by considering, respectively, the small-$s$ and large-$s$ behavior of $\J^{\Psi}(s)$. The details of the analysis will then depend on whether or not the density $\psi(\ell)$ is itself heavy-tailed.
 
 \subsection{Asymptotics of the FPT density for a gamma distribution $\psi$}
\begin{figure}[b!]
\raggedleft
\includegraphics[width=10cm]{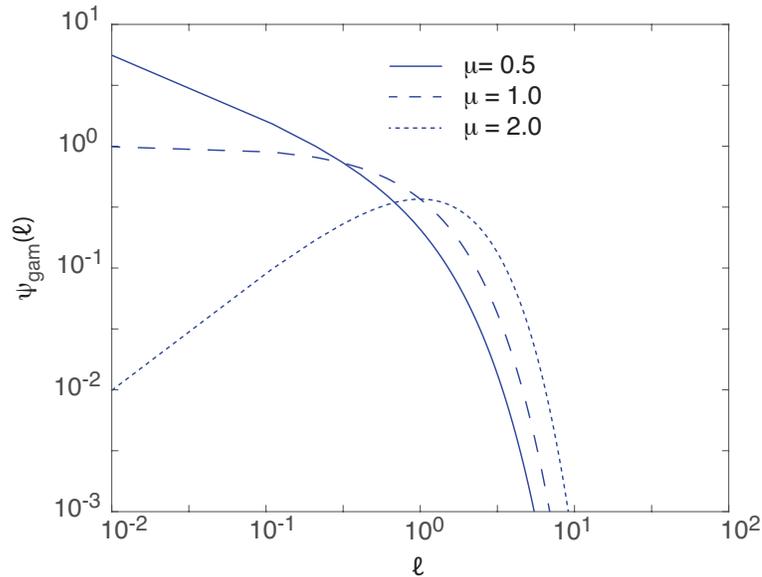} 
\caption{Plots of the probability density $\psi_{\rm gam}(\ell)$ as a function of the local time threshold for the gamma distribution with $\kappa=1$ and various values of $\mu$.}
\label{fig1}
\end{figure}

  One example of a non-exponential density with finite moments at all orders is the gamma distribution, see Fig. \ref{fig1}, 
\begin{equation}
\label{psigam}
\psi_{\rm gam}(\ell)=\frac{\kappa(\kappa \ell)^{\mu-1}\e^{-\kappa \ell}}{\Gamma(\mu)},\quad \widetilde{\psi}_{\rm gam}(z)=\left (\frac{\kappa}{\kappa+z}\right )^{\mu} 
\end{equation}
for positive constants $\kappa,\mu$ with $\mu \neq 1$, where $\Gamma(\mu)$ is the gamma function 
\begin{equation}
\Gamma(\mu)=\int_0^{\infty}\e^{-t}t^{\mu-1}dt,\ \mu >0.
\end{equation}
(The case $\mu=1$ corresponds to the exponential distribution with constant reactivity $\kappa$.) 
It can be seen from Fig. \ref{fig1} that the probability of small values of the local time threshold $\widehat{\ell}$ can be decreased relative to an exponential distribution by taking $\mu >1$. This could represent a reactive surface that is initially inactive, but becomes more activated as the number of particle-surface encounters increases, ultimately approaching a constant level of reactivity. On the other hand, the probability of small values of the local time threshold $\widehat{\ell}$ is increased when $\mu < 1$. Now the surface is initially highly reactive, but reduces to a lower constant level after a sufficient number of particle-surface encounters. 

\subsubsection{Long-time behavior.}
The Laplace transform of the gamma distribution, $\widetilde{\psi}_{\rm gam}(z)$, is analytic at $z=0$, that is, $d^m\widetilde{\psi}_{\rm gam}(z)/dz^m|_{z=0} < \infty$ for all $m\geq 0$. It follows that all moments of the gamma distribution are also finite, since
\begin{equation}
 \fl \E[\ell^m] :=\int_0^{\infty}\ell^m\psi_{\rm gam}(\ell)d\ell=\left .\left (-\frac{d}{dz}\right )^m \widetilde{\psi}_{\rm gam}(z)\right |_{z=0}<\infty \mbox { for all } m\geq 0.
 \end{equation}
 For example, the first and second moments are
\begin{equation}
\E[\ell] =\frac{\mu}{\kappa},\quad \E[\ell^2]=\frac{\mu(\mu+1)}{\kappa^2}.
\end{equation}
Analyticity of $\widetilde{\psi}_{\rm gam}$ implies that $\J^{\Psi}(s)$ is an even, analytic function of $\sqrt{s^{\alpha}/K_{\alpha}}$ for small $s$. Hence, Taylor expanding 
$ \J^{\Psi}(s)$ about $s=0$ generates a power series in $s^{\alpha}/K_{\alpha}$. To leading order we have
\begin{eqnarray}
\fl  \J^{\Psi}(s)&=& \left (1-s^{\alpha}\frac{L^2-(L-x_0)^2}{2K_{\alpha}}\right )
\left (\widetilde{\psi}_{\rm gam}(0)+\frac{s^{\alpha}L}{K_{\alpha}}\widetilde{\psi}_{\rm gam}'(0)\right )+h.o.t.  \end{eqnarray}
where $\widetilde{\psi}_{\rm gam}(0)=1$ and $-\widetilde{\psi}_{\rm gam}'(0)=\E[\ell]$.
Clearly
\begin{equation}
 \J^{\Psi}(0)\equiv \int_0^{\infty} J^{\Psi}(t)dt=1.
 \end{equation}
 A well-known result from the theory of Laplace transforms is that the large-$t$ behavior of a function $f(t)$ with $\widetilde{f}(s)\sim 1- s^{\alpha}$ for $0<\alpha <1$ and small $s$ takes the form $f(t)\sim t^{-\alpha-1}$. More precisely, 
 \begin{equation}
 J^{\Psi}(t)\sim   \frac{1}{|\Gamma(-\alpha)|t^{\alpha+1}}\left [ \frac{L^2-(L-x_0)^2+2L\E[\ell]}{2K_{\alpha}} \right ],\ t\rightarrow \infty.
 \end{equation}

We recognize the expression inside the square brackets, after replacing $K_{\alpha}$ with the diffusivity $D$, as the MFPT for adsorption at $x=0$ in the case of normal diffusion \cite{Bressloff22}. The term $(L^2-(L-x_0) ^2)/2D$ is the classical result for a totally absorbing boundary at $x=0$ and a reflecting boundary at $x=L$. The additional terms $\E[\ell]L/D$ is the contribution from paths that make one or more excursions from the boundary at $x=0$ back into the bulk domain before the particle is finally absorbed.
 In order to understand this result and to include higher-order terms in the small-$s$ expansion of $\widetilde{J}^{\Psi}(s)$, consider the corresponding flux for normal diffusion, which we write as
 \begin{equation}
 \J(s)=A(\sqrt{s/D})\widetilde{\psi}_{\rm gam}(\Gamma(\sqrt{s/D})),\quad A(y)=\frac{\cosh(y(L-x_0))}{\cosh(yL)}.
 \end{equation}
 Since both $A(y)$ and $\Gamma(y)$ are even functions of $y$, it follows that both sides can be expanded in integer powers of $s$. First,
 \begin{equation}
 \label{Jexp}
 \J(s)=\sum_{n=0}^{\infty} \frac{\J^{(n)}(0) s^n}{n!} =\sum_{n=0}^{\infty}(-1)^n \frac{\E[\calT^n]s^n}{n!},
 \end{equation}
 where $\tau_{n}^{\Psi}:=\E[\calT^n]=(-1)^n\J^{(n)}(0)$ is the $n$-th moment of the FPT density for normal diffusion in $[0,L]$. On the other hand,
 \begin{eqnarray}
 \label{Aexp}
 &A(\sqrt{s/D})\widetilde{\psi}_{\rm gam}(\Gamma(\sqrt{s/D}))\nonumber \\
&= \sum_{n=0}^{\infty} \frac{A^{(2n)}(0)}{n!}\left (\frac{s}{D}\right )^n\widetilde{\psi}_{\rm gam}\left (\sum_{k=1}^{\infty} \frac{\Gamma^{(2k)}(0)}{(2k)!}s^k/D^k\right )\\
&= \sum_{n=0}^{\infty} \frac{A^{(2n)}(0)}{(2n)!}\left (\frac{s}{D}\right )^n
 \sum_{m=0}^{\infty} \frac{\widetilde{\psi}_{\rm gam}^{(m)}(0)}{m!} \left (\sum_{k=1}^{\infty}\frac{\Gamma^{(2k)}(0)}{(2k)!}\left (\frac{s}{D}\right )^k\right )^m. \nonumber 
 \end{eqnarray}
 Note that $A(0)=1$ and $\Gamma(0)=0$. 
 Equating the power series expansions of equations (\ref{Jexp}) and (\ref{Aexp}), assuming that they are uniformly convergent for sufficiently small $s$, we find that
 \numparts
 \begin{eqnarray}
 \fl \tau_1^{\Psi}&=-\frac{A^{(2)}(0)}{2D} -\widetilde{\psi}_{\rm gam}'(0)\frac{\Gamma^{(2)}(0)}{2D}=\frac{L^2-(L-x_0)^2+2L\E[\ell]}{2D} \\
 \fl \frac{\tau_2^{\Psi}}{2}&=\frac{A^{(4)}(0)}{4!D^2}+\frac{A^{(2)}(0)\widetilde{\psi}_{\rm gam}'(0)\Gamma^{(2)}(0)}{2D^2}+\frac{\widetilde{\psi}_{\rm gam}''(0)}{8D^2}[\Gamma^{(2)}(0)]^2+\frac{\widetilde{\psi}_{\rm gam}'(0)\Gamma^{(4)}(0)}{24D^2}\nonumber \\
\fl  &=\frac{5L^4+(L-x_0)^4 -6L^2(L-x_0)^2}{4!D^2}+\frac{[L^2-(L-x_0)^2]L\E[\ell]}{D^2}\nonumber \\
\fl &\quad +\frac{\E[\ell^2]L^2}{2D^2}+\frac{L^3\E[\ell]}{3D^2}.
 \end{eqnarray}
 and the general result for all $n$ is of the form
 \begin{eqnarray}
 \fl \frac{\tau_n^{\psi}}{n!}&=\frac{(-1)^n}{D^n}\sum_{0\leq i\leq n}\, \sum_{0\leq j\leq n}\, \sum_{0\leq k\leq n}\delta_{jk,n-i}c_{ijk}^{(n)}A^{(2i)}(0)\left [\Gamma^{(2j)}(0)\right ]^k\widetilde{\psi}_{\rm gam}^{(k)}(0)
 \end{eqnarray}
 for constants $c_{ijk}^{(n)}$.
\endnumparts
Given the explicit expressions for $A(y)$ and $\Gamma(y)$ we find that
\numparts
 \begin{eqnarray}
 \fl \tau_1^{\Psi}&=\frac{L^2-(L-x_0)^2+2L\E[\ell]}{2D} \\
 \fl \frac{\tau_2^{\Psi}}{2}&=\frac{5L^4+(L-x_0)^4 -6L^2(L-x_0)^2}{4!D^2}+\frac{[L^2-(L-x_0)^2]L\E[\ell]}{D^2}\nonumber \\
\fl &\quad +\frac{\E[\ell^2]L^2}{2D^2}+\frac{L^3\E[\ell]}{3D^2}.
 \end{eqnarray}
\endnumparts
Finally, returning to the case of fractional diffusion ($0<\alpha < 1$), it follows that for small $s$
\begin{eqnarray}
\fl  \J^{\Psi}(s) &=  \sum_{n=0}^{\infty} \frac{A^{(2n)}(0)}{(2n)!}\left (\frac{s^{\alpha}}{K^{\alpha}}\right )^n
 \sum_{m=0}^{\infty} \frac{\widetilde{\psi}_{\rm gam}^{(m)}(0)}{m!} \left (\sum_{k=1}^{\infty}\frac{\Gamma^{(2k)}(0)}{(2k)!} \left (\frac{s^{\alpha}}{K^{\alpha}}\right )^k\right )^m\nonumber  \\
\fl  &=1+\sum_{n=1}^{\infty} (-1)^n\tau_n^{\psi}\left (\frac{D}{K^{\alpha}}\right )^{n}s^{n\alpha-1}.
\label{AAexp}
 \end{eqnarray}
 We thus obtain the large-$t$ approximation
 \begin{equation}
 J^{\Psi}(t)\sim \sum_{n=1}^{\infty} \frac{(-1)^n}{\Gamma(-\alpha n)n!}\left (\frac{D}{K^{\alpha}}\right )^{n}\frac{\tau_n^{\Psi} }{t^{n\alpha+1}},\quad t\rightarrow \infty.
 \label{res2}
\end{equation}
Equation (\ref{res2}) is the encounter-based generalization of the result obtained in Ref. \cite{Grebenkov10} for a constant rate of adsorption, that is, for an exponential distribution $\psi$. Note that $\Gamma(-\alpha)<0$ for $0<\alpha <1$ so that the leading-order term is positive. Moreover, if $\alpha $ is a rational number, $\alpha=p/q$ with $p,q$ having no common divisor other than unity, then all terms in the sum for which $n$ is an integer multiple of $q$ vanish. This follows from the fact that $\Gamma(-m)=\infty$ for all integers $m\geq 1$.

\begin{figure}[t!]
\raggedleft
\includegraphics[width=10cm]{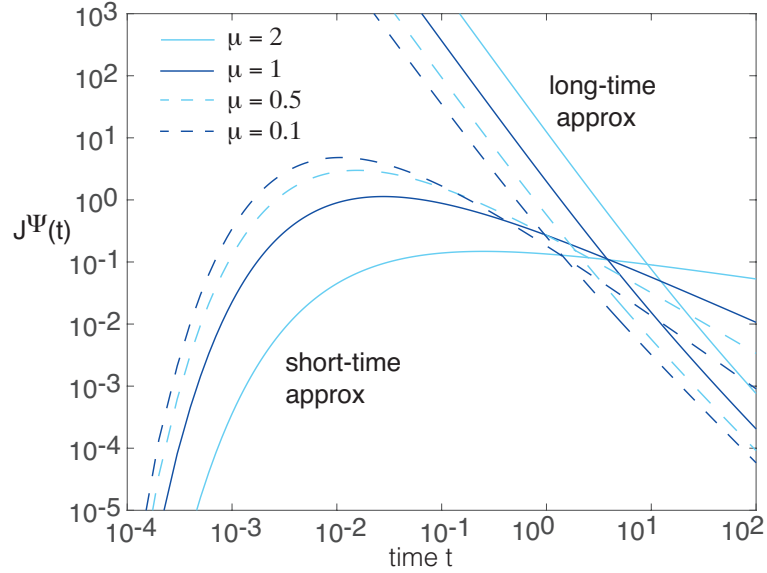}
\caption{Fractional diffusion on the interval $[0,L]$ with a partially absorbing boundary  at $x=0$ and a totally reflecting boundary at $x=L$. We plot the first two-terms in the asymptotic expansion (\ref{res2}) of the FPT density $J^{\Psi}(t)$ for the gamma distribution (\ref{psigam}) with $\E[\ell]=\mu/\kappa=1$ and various values of $\mu$. We also show the corresponding short-time approximation (\ref{res3}). Other parameters are $\alpha=2/3$, $K_{\alpha}=1$, $L=1$, and $x_0=0.75$.}
\label{fig2}
\end{figure}

\begin{figure}[t!]
\raggedleft
\includegraphics[width=10cm]{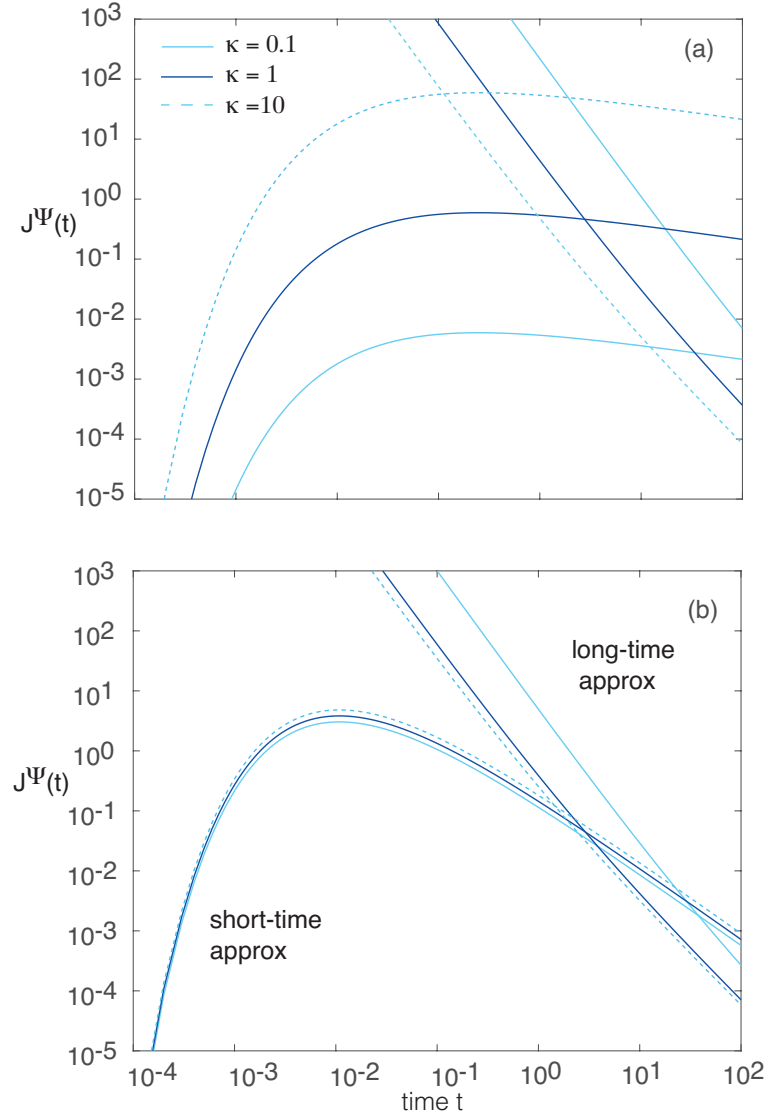}
\caption{Same as Fig. \ref{fig2} except that $J^{\Psi}(t)$ is plotted as a function of $t$ for various values of $\kappa$ with (a) $\mu=2$ and (b) $\mu=0.1$.}
\label{fig3}
\end{figure}

\subsubsection{Short-time behavior.} The small-$t$ asymptotics of $J^{\psi}(t)$ can be extracted from the large-$s$ behavior of $\J^{\Psi}(s)$. From equation (\ref{JagPsi}), we see that
\begin{eqnarray}
\J^{\Psi}(s) \sim \exp\left (-\sqrt{s^{\alpha}/K_{\alpha}} x_0)\right )\widetilde{\psi}_{\rm gam}  (\ \sqrt{s^{\alpha}/K_{\alpha}}),\quad s\rightarrow \infty.
 \end{eqnarray}
 In the case of the gamma distribution,
 \begin{eqnarray}
\J^{\Psi}(s) \sim \kappa^{\mu}\exp\left (-\sqrt{s^{\alpha}/K_{\alpha}} x_0\right )\left (\frac{s^{\alpha}}{K_{\alpha}}\right )^{-\mu/2},\quad s\rightarrow \infty.
 \end{eqnarray}
Using the approximation \cite{Grebenkov10},
\begin{eqnarray}
{\mathcal L}[s^{\beta} \e^{-\sqrt{s^{\alpha}/K_{\alpha}} x_0}]&\approx \frac{1}{t^{\beta+1}}\frac{\alpha^{\beta+1/2}}{\sqrt{\pi(2-\alpha)}} \left (\frac{x_0^2\alpha^{\alpha}}{4K_{\alpha}t^{\alpha}}\right )^{(\beta+1/2)/(2-\alpha)}\nonumber \\
&\quad \times \exp\left [-(2-\alpha)\left (\frac{x_0^2\alpha^{\alpha}}{4K_{\alpha}t^{\alpha}}\right )^{1/(2-\alpha)}\right ],
\label{stLT}
\end{eqnarray}
and setting $\beta=-\alpha \mu /2$, we obtain the leading order result 
\begin{eqnarray}
\fl J^{\Psi}(t) &\sim \frac{[\kappa\sqrt{K_{\alpha}}]^{\mu}}{t^{1-\mu \alpha/2}}\frac{[\alpha \omega(t)])^{(1-\mu \alpha)/2}}{\sqrt{\pi(2-\alpha)}} \e^{-(2-\alpha)\omega(t)},\quad \omega(t) := \left (\frac{x_0^2\alpha^{\alpha}}{4K_{\alpha}t^{\alpha}}\right )^{1/(2-\alpha)}.
\label{res3}
\end{eqnarray}

In Fig. \ref{fig2} we plot the short-term approximation (\ref{res3}) and the first two terms in the long-time approximation (\ref{res2}) of the FPT density for the gamma distribution and various values of $\mu$ and a fixed mean $\E[\ell]=\mu/\kappa$. It can be see that increasing $\mu$ increases the FPT at large times but reduces it at small times. This is consistent with the switch in the $\mu$-dependent ordering of the gamma distribution curves shown in Fig. \ref{fig1} from small values of $\ell$ to large values of $\ell$. Moreover, the log-log plots indicate that the long-term decay in $J^{\Psi}(t)$ is algebraic. Note that we focus on values of $\alpha $ in the range $1/2\leq \alpha < 1$, since more terms in the asymptotic expansion are needed for a given $t$ as $\alpha \rightarrow 0$. In Fig. \ref{fig3} we show corresponding plots of $J^{\Psi}(t)$ for fixed $\mu$ and different values of $\kappa$. This shows that the short-time approximation is insensitive to changes in $\kappa$ for small $\mu$. ( In Ref. \cite{Grebenkov10} it is shown that the short-time and long-time asymptotics agree very well with the numerical solution of the full FPT density.) 

\subsection{Asymptotics of the FPT density for a heavy-tailed distribution $\psi$} 
\begin{figure}[t!]
\raggedleft
\includegraphics[width=10cm]{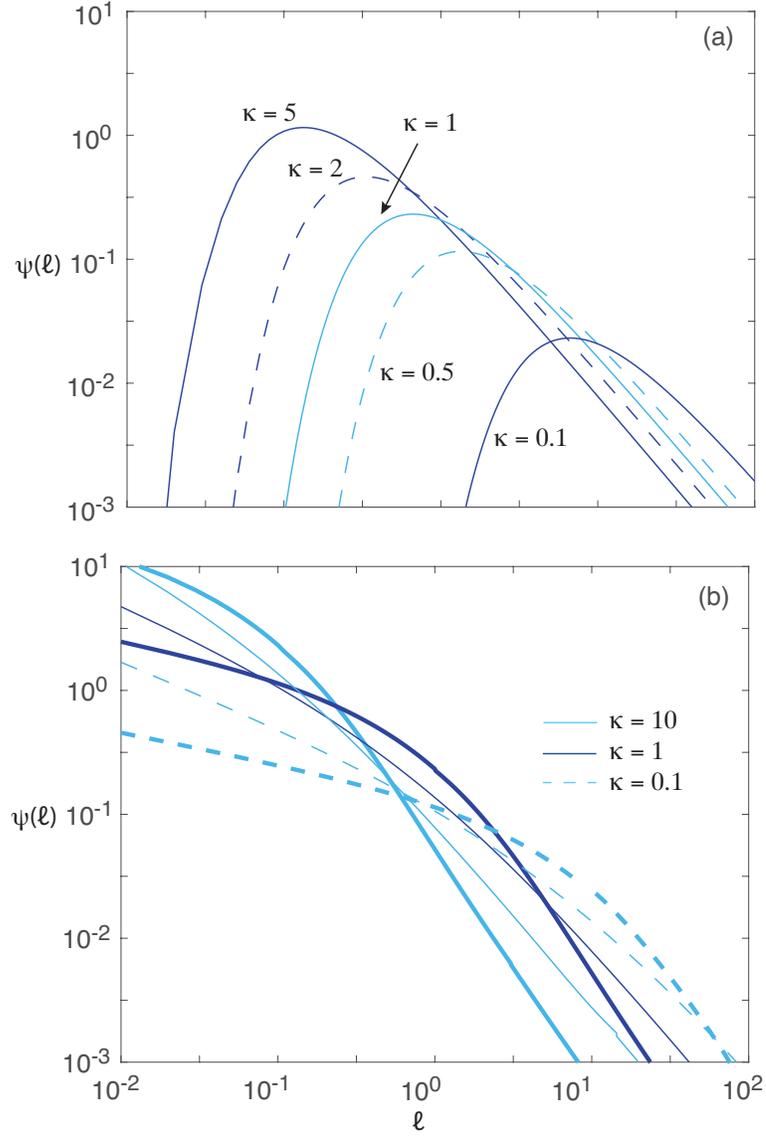} 
\caption{Two examples of a heavy-tailed distribution $\psi(\ell)$. (a) One-side Levy distribution for different values of $\kappa$. (b) Mittag-Leffler distribution for different values of $\kappa$ with $\mu=0.5 $ (thin curves) and $\mu=0.75$ (thick curves).}
\label{fig4}
\end{figure}

The Taylor expansion of $\J^{\Psi}(s)$ about $s=0$ in powers of $s^{\alpha}/K_{\alpha}$ breaks down when the local time threshold density is heavy-tailed. In particular, equations (\ref{AAexp}) and (\ref{res2}) no longer hold. However, it is still possible to perform a small-$s$ expansion for specific choices of $\widetilde{\psi}(s)$. For the sake of illustration, we consider two different examples of heavy-tailed distributions as illustrated in Fig. \ref{fig4}. (For a more extensive list, see Ref. \cite{Grebenkov20}.)

\begin{enumerate}

\item One-sided L\'evy-Smirnov distribution
 \begin{equation}
 \psi_{\rm ls}(\ell)=\kappa \frac{\e^{-1/(\kappa \ell)}}{\sqrt{\pi} (\kappa \ell)^{3/2}},\quad \widetilde{\psi}_{\rm ls}(z)=\e^{-2\sqrt{z/\kappa}}.
 \end{equation}
 This could represent a surface that has an optimal range of reactivity \cite{Grebenkov20}. Substituting for $\widetilde{\psi}_{\rm ls}$ in equation (\ref{JagPsi}) gives
 \begin{eqnarray}
  \fl  \J^{\Psi}(s) = \frac{\cosh(\sqrt{s^{\alpha}/K_{\alpha}}(L-x_0))}{\cosh(\sqrt{s^{\alpha}/K_{\alpha}}L)}\exp\left (-2\sqrt{\frac{\Gamma(\sqrt{s^{\alpha}/K_{\alpha}})}{\kappa}}\right ).
 \end{eqnarray}
 For small $s$, we have the approximation
 \begin{eqnarray}
  \fl  \J^{\Psi}(s) \approx \left (1-s^{\alpha}\frac{L^2-(L-x_0)^2}{2K_{\alpha}}\right )\left (1-2\sqrt{\frac{s^{\alpha}L}{\kappa K_{\alpha}}}\right )= 1-2\sqrt{\frac{s^{\alpha}L}{\kappa K_{\alpha}}}+ h.o.t.
 \end{eqnarray}
 Hence, the leading-order large-$t$ approximation of $J^{\Psi}(t)$ is
 \begin{equation}
 J^{\Psi}(t)\sim \frac{2}{|\Gamma(-\alpha/2)|t^{\alpha/2+1}}\sqrt{\frac{L}{\kappa K_{\alpha}}},\quad t \rightarrow \infty .
 \end{equation}
The leading order power law $t^{-\alpha/2}$ has contributions from two distinct anomalous processes. The first is subdiffusion within the bulk domain, which generates the factor $t^{-\alpha}$, whereas the additional square-root is a consequence of the heavy-tailed L\'evy distribution $\psi_{\rm ls}(\ell)$ that determines adsorption at $x=0$. Note that the short-term contribution to the FPT density is negligible due to inactivity of the boundary for small thresholds $\widehat{\ell}$, see Fig. \ref{fig4}(a).

\item Mittag-Leffler distribution
\begin{equation}
 \psi_{\rm ml}(\ell)=-E_{\mu,0}(-(\kappa \ell)^{\mu})/\ell, \quad E_{\mu,0}(z)=\sum_{k=0}^{\infty} \frac{z^k}{\Gamma(\mu k)}
 \end{equation}
for $0<\mu<1$. The corresponding Laplace transform is
\begin{equation}
 \widetilde{\psi}_{\rm ml}(z)=\frac{\kappa^{\mu}}{\kappa^{\mu}+z^{\mu}}.
 \end{equation}
 Substituting for $\widetilde{\psi}_{\rm ml}$ in equation (\ref{JagPsi}) gives
 \begin{eqnarray}
 \J^{\Psi}(s) = \frac{\cosh(\sqrt{s^{\alpha}/K_{\alpha}}(L-x_0))}{\cosh(\sqrt{s^{\alpha}/K_{\alpha}}L)}\frac{\kappa^{\mu}}{\kappa^{\mu}+\Gamma(\sqrt{s^{\alpha}/K_{\alpha}})^{\mu}}.  
 \end{eqnarray}
 For small $s$, we have the approximation
 \begin{eqnarray}
 &\J^{\Psi}(s) \approx \left (1-s^{\alpha}\frac{L^2-(L-x_0)^2}{2K_{\alpha}}\right )\left (1-\left (\frac{s^{\alpha}L}{\kappa K_{\alpha}}\right )^{\mu}\right )\nonumber \\
  &= 1-s^{\alpha}\frac{L^2-(L-x_0)^2}{2K_{\alpha}}-\left (\frac{s^{\alpha}L}{\kappa K_{\alpha}}\right )^{\mu}+ h.o.t.
 \end{eqnarray}
 Since $0<\mu < 1$, it follows that the $s^{\alpha \mu}$ term dominates for small $s$.
Hence, the leading-order large-$t$ approximation of $J^{\Psi}(t)$ is
 \begin{equation}
 J^{\Psi}(t)\sim \frac{1}{t^{\alpha \mu+1}}\left (\frac{ L}{\kappa K_{\alpha}}\right )^{\mu},\quad t \rightarrow \infty .
 \end{equation}
 Again the characteristic power law $t^{-\alpha \mu}$ has contributions from anomalous bulk diffusion and surface adsorption. On the other hand, the short-time behavior is identical to a gamma distribution with the parameters $(\mu,k)$.

\end{enumerate}

\section{Higher spatial dimensions}

\begin{figure}[b!]
\raggedleft
\includegraphics[width=10cm]{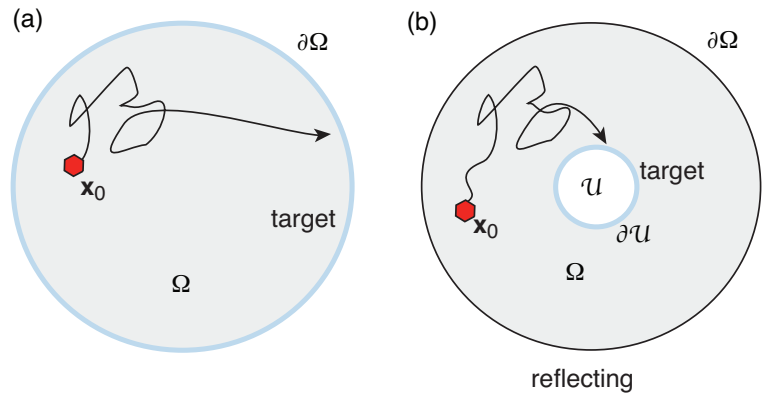} 
\caption{Higher-dimensional (sub)diffusion to a target surface. (a) Target surface is the exterior boundary $\partial \Omega$ of a simply-connected bounded domain $\Omega$. (b) Target surface is an interior boundary $\partial \calU$ with $\calU\subset \Omega$ and $\partial \Omega$ totally reflecting.}
\label{fig5}
\end{figure}

Consider a $d$-dimensional version of the local time propagator equations (\ref{fracPa}) and (\ref{fracPb}) for fractional diffusion. Suppose that a particle diffuses in a bounded, simply connected domain $\Omega\subset \R^d$ with a smooth boundary $\partial \Omega$, see Fig. \ref{fig5}(a). Introduce the boundary local time
\begin{equation}
\label{dloc}
\fl \ell(t)=\lim_{\epsilon\rightarrow 0} \frac{D}{\epsilon} \int_0^tH(\epsilon-\mbox{dist}(\X(\tau),\partial \Omega^-))d\tau=D\int_0^{t} \left [\int_{\partial \Omega} \delta(\X(\tau)-\y)d\y\right ]d\tau,
\end{equation}
where $H(x)$ is the Heaviside function and $\mbox{dist}(\X_{\tau},\partial \Omega)$ denotes the shortest Euclidean distance of $X(\tau)$ from the boundary $\partial \Omega$. The $d$-dimensional propagator equation takes the form
\numparts
\label{qdiffloc}
\begin{eqnarray}
	&\frac{\partial P(\x,\ell,t)}{\partial t} =K_{\alpha}{\mathcal D}_t^{1-\alpha}\nabla^2 P(\x,\ell,t) \mbox{ for } \x \in  \Omega,\\
&-\nabla P(\x,\ell,t) \cdot \n=\frac{\partial P}{\partial \ell}(\y,\ell,t)  +\delta(\ell)P(\y,0,t) \mbox{ for } \y\in \partial \Omega,
\label{qdiffloc2}
	\end{eqnarray}
	\endnumparts	
	where $\n$ is the outward unit normal at a point on the surface $\partial \Omega$
Again this can be derived by taking a continuum limit of a heavy-tailed CTRW on a $d$-dimensional regular lattice. Using analogous arguments to the 1D case, we introduce the FPT (\ref{1DTell}) for adsorption. Given the threshold distribution $\Psi(\ell)$, the marginal probability density for particle position can be written as
\begin{eqnarray}
     \label{int}
  p^{\Psi} (\x,t)&=\int_0^{\infty} \Psi(\ell){\mathcal L}_{\ell}^{-1}[G(\x,z,t)]d\ell ,\quad \x\in \Omega,
  \end{eqnarray}
  with
\numparts
\label{qLTa}
\begin{eqnarray}
	&K_{\alpha}{\mathcal D}_t^{1-\alpha}\nabla^2 G(\x,\ell,t) -sG(\x,\ell,t)=-\delta(\x-\x_0)\mbox{ for } \x,\x_0 \in  \Omega,\\
&-\nabla G(\y,z,t) \cdot \n=zG(\y,z,t) \mbox{ for } \y\in \partial \Omega.
\label{qTb}
	\end{eqnarray}
	\endnumparts	
Similarly, the total flux through $\partial \Omega$ is
\begin{equation}
\label{job}
J^{\Psi}(t)=K_{\alpha}\int_0^{\infty} \psi(\ell)  \left [\int_{\partial \Omega}{\mathcal D}_t^{1-\alpha}  {\mathcal L}_{\ell}^{-1}[G(\y,z,t)] d\y \right ]d\ell .
\end{equation}

A crucial observation is that the small-$s$ series expansion (\ref{AAexp}) and its corresponding large-$t$ expansion (\ref{res2}) still hold for the gamma distribution $\psi_{\rm gam}$, except that $\tau_n^{\Psi}$ are now the moments of the FPT density for normal diffusion in $\Omega\subset \R^d$. Hence, calculating the large-$t$ asymptotics reduces to finding the higher-dimensional analogs of the functions $\Gamma(y)$ and $A(y)$ of equations (\ref{JagPsi}) and (\ref{gamy}). One configuration where this can be achieved is for a $d$-dimensional sphere. Suppose that $\Omega =\{\x\in \R^d\,|\, 0\leq  |\x| <R\}$ and thus $\partial \Omega =\{\x\in \R^d\,|\,  |\x| =R\}$, where $R$ is the radius of the sphere.
We assume that the initial distribution of the particle is spherically symmetric, that is, $G(\x,z,0)=\delta(|\x|-r_0)/\Omega_d r_0^{d-1}$, where $\Omega_d$ is the surface area of the unit sphere in $\R^d$ and $0<r_0<R$. This allows us to exploit spherical symmetry by setting $G(\x,z,t)=G(r,z,t)$ with $r=|\x|$.
The Laplace transformed propagator $\G(r,z,s)$ satisfies the modified Helmholtz equation equation
\numparts
\begin{eqnarray}
 &K_{\alpha}s^{1-\alpha} \left [\frac{\partial^2\G(r,z,s)}{\partial r^2} + D\frac{d - 1}{r}\frac{\partial \G(r,z,s)}{\partial r}\right ]-s\G(r,z,s)\nonumber \\
 &\hspace{3cm} =-\Gamma_d\delta(r-r_0) ,\quad R <r,
 \label{spha}\\
  &D\frac{\partial \G(r,z,s)}{\partial r}=-z \G(r,z,s) ,\quad r=R,
  \label{sphb}
\end{eqnarray}
\endnumparts
with $\Gamma_d=1/(\Omega_dr_0^{d-1})$.
Equations of the form (\ref{spha}) and (\ref{sphb}) can be solved in terms of modified Bessel functions \cite{Yuste07,Grebenkov10,Bressloff23}. In particular,  
\begin{eqnarray}
    \G(R, z,s) = {A}(z,s)\sqrt{s^{\alpha}/K_{\alpha}}{F}(\sqrt{s^{\alpha}/K_{\alpha}} r_0) ,\quad   0< r,r_0< R,\nonumber \\
    \label{qqir}
\end{eqnarray}
with 
\begin{equation} 
\label{Fx}
{F}(x)=\left \{ \begin{array}{cc}\cosh( x) & d=1,\\  I_0(x) & d=2,  \\  \frac{\displaystyle \sinh(  x)}{\displaystyle x} & d=3. \end{array} \right .
\end{equation}
and
\begin{eqnarray}
\label{AAbar}
\fl  {A}(z,s)=\frac{1}{s}\frac{1}{\Omega_dR^{d-1}} \frac{\sqrt{s^{\alpha}/K_{\alpha}}}{\sqrt{s^{\alpha}/K_{\alpha}}F'(\sqrt{s^{\alpha}/K_{\alpha}} R) +zF (\sqrt{s^{\alpha}/K_{\alpha} }R) }.
\end{eqnarray}
Substituting into the Laplace transform of equation (\ref{job}) and imposing spherical symmetry yields
\begin{eqnarray}
\fl \J^{\Psi}(s)&=\int_0^{\infty}\psi(\ell) {\mathcal L}_{\ell}^{-1} \left [\frac{{F}(\sqrt{s^{\alpha}/K_{\alpha}} r_0)}{\sqrt{s^{\alpha}/K_{\alpha}}F'(\sqrt{s^{\alpha}/K_{\alpha}} R) +zF (\sqrt{s^{\alpha}/K_{\alpha} }R) }\right ]d\ell\\
\fl &=\frac{F(\sqrt{s^{\alpha}/K_{\alpha}} r_0)}{F (\sqrt{s^{\alpha}/K_{\alpha} }R) }\widetilde{\psi}
 (\Gamma(\sqrt{s^{\alpha}/K_{\alpha}})),
\end{eqnarray}
with
\begin{equation}
\Gamma(y)= \frac{y F'(y R) }{F(yR)}.
\end{equation}

Note that the 1D version of the configuration shown in Fig. \ref{fig5}(a) has a reflecting boundary at $x=0$ and a partially absorbing boundary at $x=R$. Hence, we recover the 1D result on setting $F(yR)=\cosh(yL)$ and $F(yr_0)=F(y[L-x_0)$. Another configuration that reduces to the equivalent 1D problem is shown in Fig. \ref{fig5}(b). This consists of a pair of concentric spheres of radii $R_0$ and $R$ with $R_0<R$. Now there is a reflecting boundary at $x=R$ and a partially absorbing boundary at $x=R-R_0$. The higher-dimensional version can be analyzed along similar lines to the previous case by exploiting spherical symmetry, see also Ref. \cite{PCBII}.

\section{Conclusion} One of the characteristic features of encounter-based models is that the stochastic process of surface adsorption is separated from the stochastic dynamics in the bulk. This allows one to incorporate non-Markovian models of adsorption that depend non-exponentially on the amount of particle-surface contact time. In this paper we exploited this feature in order to investigate how non-Markovian surface adsorption affects the long-time power-law decay of the FPT density for subdiffusion in a bounded domain $\Omega$, see Fig. \ref{fig5}(a). In a companion paper \cite{PCBII}, we consider the complementary problem of a particle diffusing in $\Omega$ with a partially absorbing interior trap, see Fig. \ref{fig5}(b). However, rather than taking $\partial \calU$ to be absorbing, we allow the particle to freely enter and exit $\calU$ until it is eventually absorbed somewhere within $\calU$. This type of scenario was previously considered in the case of normal diffusion, where the relevant particle-surface contact time is the Brownian occupation time of $\calU$ \cite{Bressloff22,Bressloff22a}. In order to incorporate subdiffusion, we derive a fractional diffusion equation for the occupation time propagator by taking the continuum limit of a corresponding heavy-tailed CTRW. We use the model to determine conditions under which the MFPT for absorption within the trap is finite, assuming that the particle diffuses normally within $\Omega \backslash \calU$ and suddiffusively with $\calU$.

\end{document}